\documentclass[english,aps,prl,preprint,showpacs]{revtex4}
\usepackage[latin1]{inputenc}
\usepackage{babel}
\usepackage{graphics}

\makeatletter

\providecommand{\LyX}{L\kern-.1667em\lower.25em\hbox{Y}\kern-.125emX\@}

\makeatother
\begin{document}

\title{Understanding the complex patterns of snow crystals}

\author{Francisco Vera}

\email{fvera@ucv.cl}

\affiliation{Universidad Cat\'{o}lica de Valpara\'{\i}so, Av. Brasil 2950, Valpara\'{\i}so,
Chile }

\begin{abstract}
We will show that the complex shapes of snow crystals can be explained
from a simple basic mechanism that is also responsible for the appearance
of many others structures in nature. We expect that this new physical
mechanism, that follows from minimizing the total stored energy, will
permit to explain most of the features of snow crystal growth.
\end{abstract}


\maketitle
Up to now, we all have admired snowflakes and other beautiful examples
of solidification patterns, but the physical mechanism responsible
for this structures has remained a mystery. In this work we will unveil
the underlying basic mechanism responsible for the spontaneous pattern
formation in the growth of crystals. We will provide an example of
evolution of a macroscopic pattern obtained using a numerical implementation
of the basic equations, starting from a single point and evolving
in a deterministic way towards a complex branched structure.

Real snow crystals often come with a six-fold symmetrical shape as
a consequence that the most typical basic form of an ice crystal is
an hexagonal prism. Other basic forms occur at very low temperatures
or very high pressures.

A typical planar dendritic ice crystal showing the six-fold symmetry
is shown in Fig. 1. Snow crystals grow from condensing water vapor
in the air, around a nucleus of dust. In their travel from cloud towards
ground, these crystals pass through various temperature regimes. If
ambient temperature is cold enough, their arms grow very rapidly,
and if temperature warms up, the arms are capped, spreading outward
slightly, until the next cold temperature regime causes them to grow
again. What we see landing at earth surface are snowflakes, formed
by snow cristals aglomerations produced at warmer zones of the atmosphere.

\begin{figure}
{\centering \resizebox*{1\columnwidth}{!}{\includegraphics{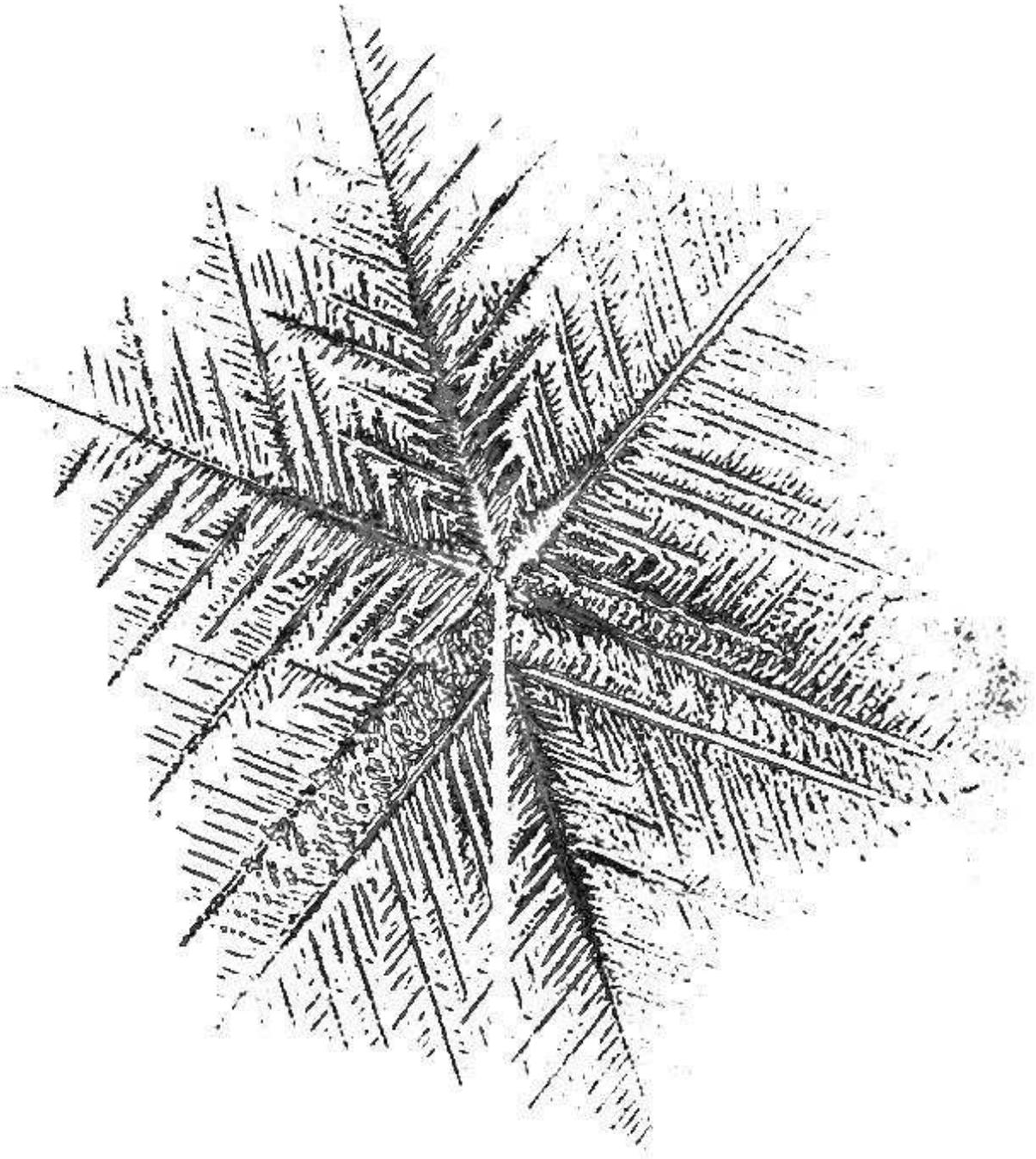}} \par}

\caption{Photograph of dendritic ice crystal grown in pure water at an undercooling
of TM - T = 2.34\protect\( ^{\circ }\protect \)C ( Original by T.
Fujioka, reproduced from Langer 1980. We altered the background of
this image).}
\end{figure}

The observation of snowflakes and these sub-millimeter snow crystals
can be traced to 1611 with the work of Johannes Kepler entitled \char`\"{}The
Six-Cornered Snowflake\char`\"{} \cite{key-2}. After that, several
authors have continued the work of observing, cataloging, taking photos,
and growing artificial snow crystals in the laboratory \cite{key-3,key-4,key-5,key-6,key-7,key-8,key-9,key-10,key-11}.
The preliminary observations together with the latest systematic observations
and experiments, have provided the basis for understanding the basic
mechanism responsible for the creation of these fascinating structures.

For the theoretical description of crystal growth, and reviews on
the vast subject of pattern formation, we will refer the reader toward
Refs. \cite{key-12,key-13,key-14,key-15,key-16}. For alternative
models of crystal growth we refer the reader toward Refs. \cite{key-17,key-18,key-19}.

The key point for understanding the physics governing crystal growth
comes from studying from an unified point of view, several apparently
different physical systems. We studied several pattern forming systems
in which the normal growth velocity is proportional to the gradient
of a bulk field which obeys a Laplace or diffusion type of equation.
Some examples are: dielectric breakdown, streamers, solidification,
crystal growth, viscous fingers, biological patterns, combustion,
etc. The origin of structures in these systems is known to be related
to the Mullins-Sekerka instability \cite{key-20,key-21}. This instability
explain why a protrusion in the interface causes the interface to
grow faster there, but the theory for the evolution of a growing protrusion
towards complex structures has remained a mystery. In this work we
will show that the underlying basic mechanism responsible for spontaneous
pattern formation, needs to consider the global evolution of the system
and not just to study the evolution of a local protrusion. It is not
true that just a protrusion grows faster, and depending on some physical
parameters, sometimes is energetically favorable for the system to
develop other branches after some size of the protrusion is reached.

In the theoretical study of the evolution of an interface, the simplest
possibility is to expand the growing interface towards regions of
maximum temperature gradient. This necessarily gives a straight line
as a consequence of the Mullins-Sekerka instability mentioned previously.
We will show that it is possible to obtain a branched structure starting
from a central seed, following a quasi-stationary deterministic treatment,
that only relies in minimizing the total energy stored in the system,
and changing locally the thermal conductivity of the medium at each
step of iteration. 

The energy stored in the temperature field can be obtained from

\begin{equation}
\label{eq1}
U=\frac{1}{2}\int k(\nabla T\cdot \nabla T)dV.
\end{equation}
By analogy with dielectric breakdown \cite{key-22}, we can study
systems with boundary conditions for the temperature T at the inner
and outer interface, by letting them to evolve towards regions of
higher U. This turn out to be equivalent to the evolution towards
a pattern that minimize the total stored energy, after a local change
in the thermal conductivity k.

\begin{figure}
{\centering \resizebox*{1\columnwidth}{!}{\includegraphics{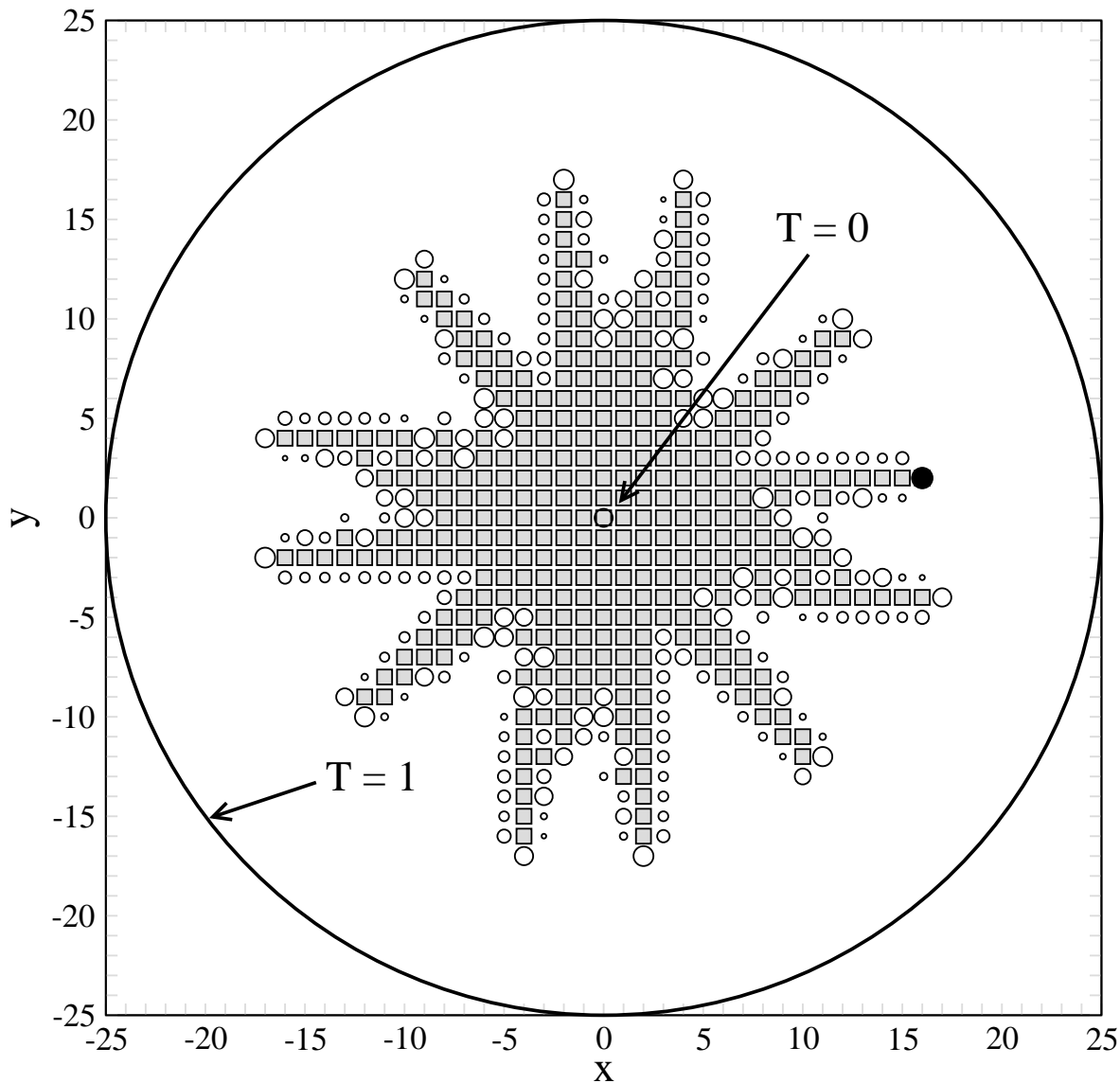}} \par}

\caption{Schematic view of the two-dimensional circular geometry used in our
calculations. The central circle represents the seed and the big circle
the outer interface. Filled boxes represent the evolving pattern (sites
where the thermal conductivity is greater). The circles show all possible
sites where the pattern can evolve, the diameter of each of these
circles represent the value for the energy U of the system in case
this site is added to the pattern. The black circle at the right of
the figure is the next step in the evolution of this pattern.}
\end{figure}

The energy U can be obtained as follows: First, impose the boundary
conditions: T= 0 in the inner seed and T= 1 in the outer interface.
Second, impose a fixed constant value for the thermal conductivity
in all the region between the outer interface and the seed. Third,
solve the Laplace equation \begin{equation}
\label{eq2}
\nabla \cdot (k\nabla T)=0,
\end{equation}
obtaining the temperature T in the region between the outer interface
and the seed. Fourth, obtain the gradients of T, and using Eq. 1,
obtain the energy U.

Now we will study the evolution of a central seed in the circular
geometry of Fig. 2. We consider a two-dimensional square lattice,
where the central point is the seed and the outer interface is modeled
as a discretized version of the big circle. The boundary conditions,
T=0 in the seed and T=1 in the outer interface, are maintained trough
all the steps in our simulation. The first step is to assign a fixed
value for the thermal conductivity of each lattice point between the
seed and the outer interface, and set the initial pattern as the central
point. The second step is to obtain the energies U of the system after
changing the thermal conductivity in one of each neighbor of the pattern
to a greater value k', these energies are compared and the neighbor
providing the bigger energy value is added to the pattern. To maintain
the model simple, pattern evolution through the diagonals are not
permitted. The pattern grows adding a site to the evolving pattern
each time the last step is repeated. The filled boxes represent the
evolving pattern (sites where the thermal conductivity is k'). The
circles show all possible sites where the pattern can evolve, the
diameter of each of these circles represent the value for the energy
U of the system in case this site is added to the pattern. The black
circle at the right of the figure shows the site giving the biggest
contribution to U, and is the next step in the evolution of this pattern.
System parameters are the same as those used for obtaining Fig. 3,
see below.

\begin{figure}
{\centering \resizebox*{1\columnwidth}{!}{\includegraphics{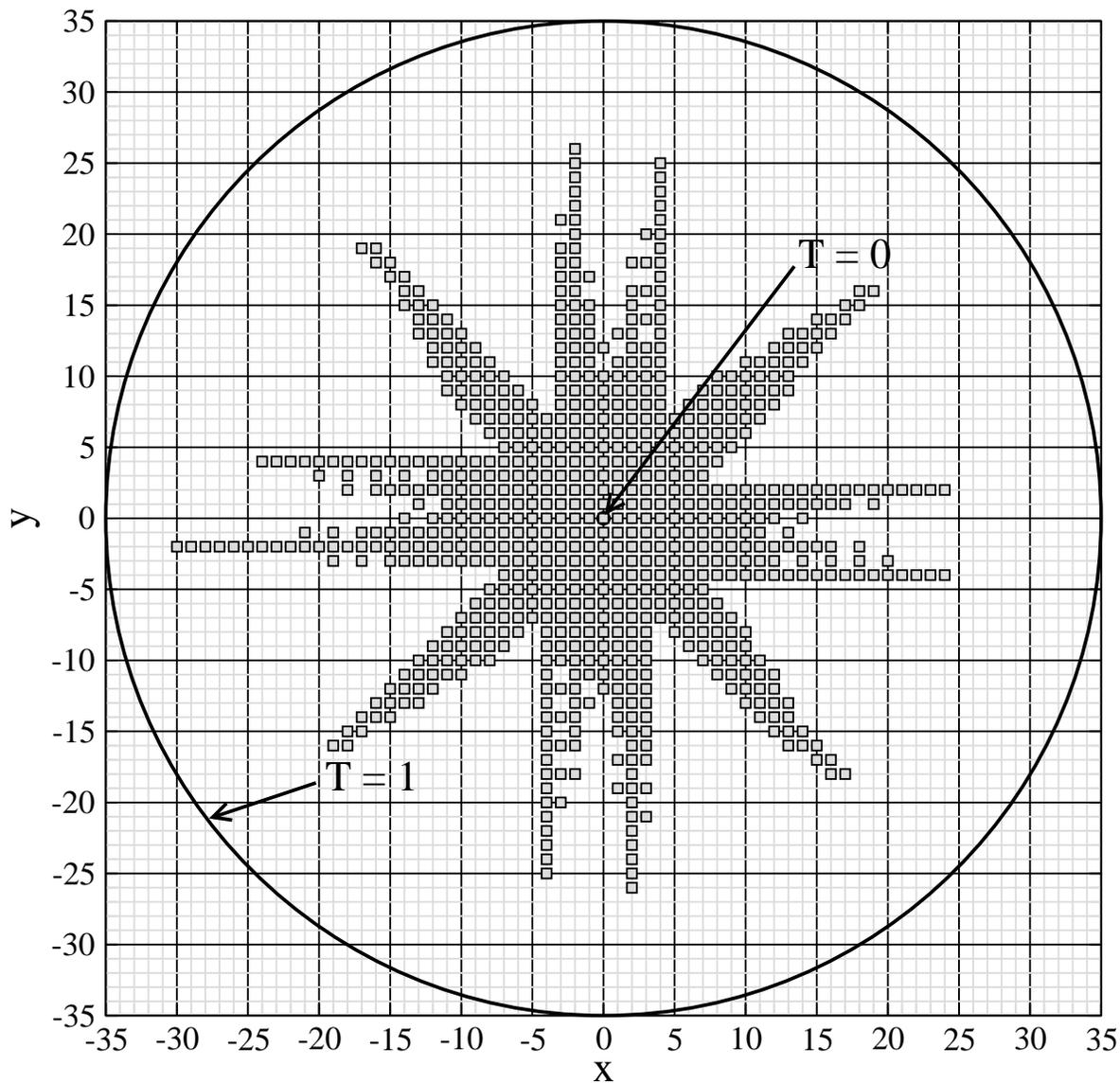}} \par}

\caption{Pattern developed for a 70 x 70 square lattice after 750 iterations
for thermal conductivity outside the channel k =2 and inside the channel
k'=6.}
\end{figure}

Fig. 3 shows the structure of the pattern for a 70x70 square lattice
after 750 iterations. For each different configuration, the numerical
solution of Eq. 2 was accepted when the numerical residual was less
than 10\textasciicircum{}\{-1\}, the values for the thermal conductivity
outside the channel was k =2 and inside the channel was k'=6. The
evolution of the pattern shows that opposite branches are not exactly
aligned, we consider this as a prediction of our model and with a
little bit of imagination it is possible to find this effect in the
experimental results shown in Fig. 1. Our example also shows that
the system develops, forming initially a central structured core,
this core supports the evolution of the main venous branches and diagonal
branches. We expect that secondary branches emerging from the main
branches will appear after some iterations, resembling the branches
shown in Fig. 1. Our simulated structure doesn't show the six-fold
symmetry of real snow crystals, this is a consequence of using a square
lattice in our numerical model. We expect that realistic six-fold
patterns will be formed in a lattice that impose the hexagonal symmetry.
We have not tested our model changing the supporting lattice, because
several months of computing time was needed for completing this example.
Repetitive calculations, or calculations using bigger lattices, would
require to implement sophisticated numerical methods \cite{key-23}.

The large computing time needed for these simulations is a consequence
of the necessity for the system to probe different possible configurations
and select the one minimizing the total stored energy. This is the
famous principle of minima action that is nicely explained in the
textbook of Richard Feynman \cite{key-24}. In normal systems, like
a mass falling towards the earth surface as a consequence of the force
of gravity, this principle leads to simple differential equations
for the motion of particles. In the case studied in this article there
are not trajectories emerging from this principle, instead there are
complex configurations of the inner interface as a result of minimizing
the total stored energy. We expect that, at least for some specific
examples, our discretized model could evolve towards a continuum theory
in the future. In either case the fundamental principles discussed
in this paper would help to explain many systems developing complex
structures.

\end{document}